\documentclass[fleqn,usenatbib]{mnras}

\usepackage{newtxtext,newtxmath}

\usepackage[T1]{fontenc}

\DeclareRobustCommand{\VAN}[3]{#2}
\let\VANthebibliography\thebibliography
\def\thebibliography{\DeclareRobustCommand{\VAN}[3]{##3}\VANthebibliography}

\usepackage{graphicx}	
\usepackage{amsmath}	

\title[Observations of SBS~0846+513]{Observations of the gamma-ray emitting narrow-line Seyfert 1, SBS~0846+513, and its host galaxy}

\author[T. S. Hamilton et al.]{
Timothy S. Hamilton,$^{1}$\thanks{E-mail: thamilton@shawnee.edu}
Marco Berton,$^{2,3}$
Sonia Ant\'{o}n,$^{4,5}$
Lorenzo Busoni,$^{6}$
Alessandro Caccianiga,$^{7}$
\newauthor
Stefano Ciroi,$^{8}$
Wolfgang G\"{a}ssler,$^{9}$
Iskren Georgiev,$^{9}$
Emilia J\"{a}rvel\"{a},$^{10}$
S. Komossa,$^{11}$
Smita Mathur,$^{12}$
\newauthor
Sebastian Rabien$^{13}$
\\
$^{1}$Department of Natural Sciences, Shawnee State University, 940 2nd St., Portsmouth, OH 45662, USA\\
$^{2}$Finnish Centre for Astronomy with ESO (FINCA), University of Turku, Quantum, Vesilinnantie 5, FI-20014 University of Turku, Finland\\
$^{3}$Aalto University Mets\"{a}hovi Radio Observatory, Mets\"{a}hovintie 114, FI-02540 Kylm\"{a}l\"{a}, Finland\\
$^{4}$CIDMA, Departmento de F\'{i}sica, Universidade de Aveiro, Campus Universit\'{a}rio de Santiago, 3810-193 Aveiro, Portugal\\
$^{5}$Instituto de Telecomunica\c c\~oes, Campus Universit\'{a}rio, 3810-193 Aveiro, Portugal\\
$^{6}$Arcetri Astrophysical Observatory, Largo Enrico Fermi 5, I-50125 Florence, Italy\\
$^{7}$INAF - Osservatorio Astronomico di Brera, Via Brera 28, 20121 Milano, Italy\\
$^{8}$Department of Physics and Astronomy, Padova University, Vicolo dell'Osservatorio 3, I-35122 Padova, Italy\\
$^{9}$Max-Planck Institut f\"{u}r Astronomie, K\"{o}nigstuhl 17, D-69117 Heidelberg, Germany\\
$^{10}$European Space Agency, European Space Astronomy Centre, C/ Bajo el Castillo s/n, 28692 Villanueva de la Ca\~{n}ada, Madrid, Spain\\
$^{11}$Max-Planck-Institut f\"{u}r Radioastronomie, Auf dem H\"{u}gel 69, D-53121 Bonn, Germany\\
$^{12}$Department of Astronomy, The Ohio State University, 140 West 18th Avenue, Columbus, OH 43210, USA; \\Center for Cosmology and AstroParticle Physics, The Ohio State University, 191 West Woodruff Avenue, Columbus, OH 43210, USA\\
$^{13}$Max-Planck Institut f\"{u}r Extraterrestrische Physik, Giessenbachstrasse 1, D-85748 Garching, Germany
}

\date{Accepted XXX. Received YYY; in original form ZZZ}

\pubyear{2020}

\begin{document}
\label{firstpage}
\pagerange{\pageref{firstpage}--\pageref{lastpage}}
\maketitle

\begin{abstract}
The gamma-ray emitting galaxy SBS~$0846+513$ has been classified as a Narrow-Line Seyfert 1 from its spectroscopy, and on that basis it was thought likely to have a small central black hole hosted in a spiral galaxy.  But very few of the gamma-ray Narrow-Line Seyfert 1s have high-resolution imaging of their hosts, so it is unknown how those expectations hold up for the gamma-emitting class.  We have observed this galaxy in the $J$-band with the Large Binocular Telescope's LUCI1 camera and the ARGOS adaptive optics system.  We estimate its black hole mass to lie between $7.70 \leq \log \frac{\text{M}}{\text{M}_\odot} \leq 8.19$, using the correlation with bulge luminosity, or $7.96 \leq \log \frac{\text{M}}{\text{M}_\odot} \leq 8.16$ using the correlation with S\'{e}rsic index, putting its mass at the high end of the Narrow Line Seyfert 1 range.  These estimates are independent of the Broad Line Region viewing geometry and avoid underestimates due to looking down the jet axis.  Its host shows evidence of a bulge + disc structure, both from two-dimensional modeling and isophote shape, in keeping with the expectations.  Mergers and interactions appear to be common among the gamma-ray Narrow-Line Seyfert 1s, and we see some circumstantial evidence for companion galaxies or disturbed features in the host.  
\end{abstract}




\section{Introduction}\label{sec:intro}

The relation that links galaxies with the black hole (BH) at their centres has been known for a while, and it is valid for non-active and active galaxies (e.g., \citealt{2000ApJ...539L...9F}). The host galaxy of an active galactic nucleus (AGN) is the closest environment with which the nucleus interacts and co-evolves \citep{2013ARA&A..51..511K}. For example, accretion on to the central BH can ionize and heat the gas, suppressing the star formation (SF) in the circumnuclear region \citep{2012Natur.485..213P}. Simultaneously, the host directly affects the nuclear activity by regulating the gas supply of the BH. A remarkable example of this feedback is the interplay between the host and the relativistic plasma jet that can be launched by the BH. For example, jets are observed digging their way through the host galaxy gas \citep{2015ASPC..499..125M}, and they can efficiently suppress the SF activity far from the nucleus \citep{2016ApJ...826...29L}. Although jets are crucial components in some AGN, they are far from understood. Why and how they are formed is still unclear, but the host galaxy likely plays a crucial role in this. In particular, powerful relativistic jets are preferentially, though not exclusively, associated with ellipticals \citep{2000ApJ...543L.111L}, and merging or interaction may be the keys that trigger the jet activity \citep{2015ApJ...806..147C}.

An important new laboratory to study the jet phenomenon is the AGN class known as narrow-line Seyfert 1 (NLS1s).  Many of the sources referred to as Narrow-Line {\it Seyfert} 1s are actually {\it quasars} under the standard definition (total magnitude $M < -23$), though this limit does not distinguish between the host and nuclear emission.  The physical arguments regarding the Narrow-Line Seyfert 1s we expect to apply to narrow-line type-1 quasars, as well, and in this discussion, we refer to them collectively as NLS1.  These objects were first classified by \citet{1985ApJ...297..166O} and are defined by the low full width at half maximum of H~$\beta$ ($<$2000 km s$^{-1}$; \citealt{1989ApJ...342..224G}). The narrowness of permitted lines is believed to be due to the low rotational velocity around a low-mass black hole ($<10^8 \, \text{M}_\odot$; \citealt{2004ApJ...606L..41G}).  Other lines of reasoning support the idea of small BH masses for NLS1, including their X-ray power spectrum distribution \citep{2018ApJ...866...69P} and the rapidity of their X-ray variability \citep{1997MNRAS.289..393B,1999MNRAS.303L..53B,2015AJ....150...23Y,2018rnls.confE..34G}.    Taking into account their luminosity, comparable to that of broad-line Seyfert 1s, this is suggestive of a high Eddington ratio \citep{1992ApJS...80..109B}. These properties have been interpreted as a sign of young age \citep{2000MNRAS.314L..17M}. If this is the case, NLS1s may be the low-$z$ analog of the progenitors of high-mass AGN, sharing properties with early quasars.  

However, it has been claimed that NLS1s are not genuinely young AGN (e.g., \citealt{2018A&A...616A..43S}).  Narrow permitted lines may also be due to the projection effects caused by a flattened broad-line region observed pole-on \citep{2008MNRAS.386L..15D}.  Were this the case, the BH masses derived from spectral line widths \citep{2013MNRAS.431..210C} would be underestimated.  Fortunately, the host galaxy can help disentangling these two scenarios.  Generally speaking, ellipticals tend to harbor more massive BH than spirals (e.g., \citealt{2000MNRAS.317..488S}), and the black hole mass correlates with properties of the bulge (e.g., \citealt{1998AJ....115.2285M,2003ApJ...589L..21M,2007ApJ...655...77G}).  So a measurement of the host galaxy type can tell us about the central black hole's mass in a way that is independent of geometry.

In 2009, $\gamma$-rays were detected from radio-loud NLS1 \citep{2009ApJ...699..976A}, and in the years since, about 19 NLS1 have been detected with the {\it Fermi Gamma-Ray Space Telescope} \citep{2018MNRAS.481.5046R,2020A&A...636L..12J}.  These have shown unambiguously that there exists a subset of NLS1 that are radio-loud and have powerful, relativistic jets, even though with their smaller black holes, they may have been expected not to \citep{2000ApJ...543L.111L}.  The $\gamma$-NLS1s are an intriguing and still poorly-understood sample.  If they are from the same broader population as the radio-quiet NLS1, then what triggers the relativistic jet?  As described by \citet{2018rnls.confE..15K}, a large number of $\gamma$-NLS1 seem to be in interacting systems.  In fact, of the sources analysed in the study, only PKS~$2004-447$ and SBS~$0846+513$ did not show evidence for interaction at the then-available resolution.  On the other hand, the majority of radio-quiet NLS1 do not show an excess of companion galaxies or evidence for mergers \citep{2001AJ....121..702K,2012AJ....143...83X}.  In general, jetted NLS1 preferably reside in denser regions than do the non-jetted NLS1 sources, at least in terms of large-scale environment \citep{2018A&A...619A..69J}. 

Perhaps the host galaxies are different between the radio-quiet NLS1 and the radio-loud (especially the $\gamma$-emitters).  But only a few studies have been dedicated to the hosts of NLS1s. \citet{2003AJ....126.1690C} studied 19 NLS1s with HST, all non-jetted at $z<0.084$, finding a majority of spirals and frequent starburst rings.  A high incidence of pseudobulges was found by \citet{2011MNRAS.417.2721O} and confirmed by a subsequent study with HST of 10 more non-jetted sources up to z = 0.164 \citep{2012ApJ...754..146M}. However, nothing can be said for non-jetted sources at higher $z$. Regarding radio-loud, jetted NLS1s ($\sim$7 per cent of the whole NLS1 population; see \citealt{2006AJ....132..531K}), four sources have been studied individually \citep{2007ApJ...658L..13Z,2008A&A...490..583A,2016ApJ...832..157K,2017MNRAS.467.3712O,2017MNRAS.469L..11D,2018MNRAS.478L..66D,2019AJ....157...48B}.  On a larger sample is that by \citet{2018A&A...619A..69J} with the Nordic Optical Telescope (NOT), but even in that case only three jetted sources were successfully resolved.  In general, the results are still unclear. Most jetted NLS1s studied so far are hosted by spirals and disturbed morphologies. The recent study by \citet{2020MNRAS.492.1450O} further supports this result.  They image 29 radio-loud NLS1s, successfully detecting 21 and fitting bulge + disc models.  However, a couple of NLS1 have been found to reside in non-interacting ellipticals \citep{2017MNRAS.469L..11D,2018MNRAS.478L..66D}. This result may support the low-inclination scenario for NLS1s, and might contradict the relation between merging/interaction and jet formation, or it could be that visible signatures of past interaction have relaxed. It is clear that more examples are needed to shed light on the host-AGN connection.

We selected SBS~$0846+513$ (RA = 8$^\text{h}$49$^\text{m}$58\fs 1, Dec = +51\degr 08\arcmin 25\farcs7), a $\gamma$-ray emitting Narrow-Line Seyfert 1\footnote{SBS~$0846+513$ has total absolute magnitude $M<-23$, so it is properly considered a quasar, but as we discussed above, we group Seyferts and quasars together in our discussions of NLS1.} at $z = 0.585$, with the goals of observing the host galaxy's morphology and determining its black hole mass in a geometrically-independent way.  It was unresolved in existing imagery and did not show clear evidence of interaction, and the new observations would put it near the high-redshift end of $\gamma$-NLS1 imaged at high resolution.  Its relatively high redshift makes any investigation with ground-based telescopes difficult, but adaptive optics is a great help.  An important requirement for adaptive optics is the presence of a nearby bright reference star that can be used to model the wave front and correct the images for atmospheric distortion. However, most jetted NLS1s are located well outside the Galactic plane, and they do not have any suitable reference star in the vicinities.  For this reason, we decide to observe SBS~$0846+513$ with the Advanced Rayleigh Guided Ground Layer Adaptive Optics System (ARGOS) on the LBT Utility Camera in the Infrared (LUCI1) camera mounted on the Large Binocular Telescope (LBT).  This innovative system can produce a laser guide star that was used as reference for our observations.

The paper is organized as follows:  The source is described in \S\ref{sec:source}.  In \S\ref{sec:observations}, we present the imaging and spectroscopic observations.  The analysis is described in \S\ref{sec:analysis}, including the galaxy modeling, isophote measurements, and black hole mass calculations.  The discussion and interpretation are done in \S\ref{sec:discussion}.

\section{SBS~0846+513}\label{sec:source}

SBS~$0846+513$ was already identified as a potential blazar by \citet{1979ApJ...230...68A}, although it was considered as a BL Lac at high redshift, and its correct distance was estimated only with the Sloan Digital Sky Survey (SDSS) \citep{2003AJ....126.2579S}.  The source was already known for its extreme flaring activity, with a brightening of about 4 magnitudes in approximately a month \citep{1979ApJ...230...68A}, which was initially attributed to gravitational lensing effects \citep{1986A&A...157..383N}.  The source was identified as an NLS1 by \citet{2006ApJS..166..128Z} and \citet{2008ApJ...685..801Y}, given its H~$\beta$ FWHM of $\sim 1800$ km/s, within the NLS1 definition.  After the launch of {\it Fermi}, it was detected at $\gamma$-rays during its first observed flare \citep{2011ATel.3452....1D,2011nlsg.confE..24F}.  The black hole mass of SBS~$0846+513$ was estimated both using the FWHM and the second-order moment of H~$\beta$, and in both cases the estimate is approximately at $\sim 3\times 10^7 \, \text{M}_\odot$, well within the typical range of NLS1s \citep{2008ApJ...685..801Y,2015A&A...575A..13F}.  At radio frequencies the source has been monitored for years, showing high amplitude variability both in flux and in polarization \citep{2014ApJ...794...93M,2015A&A...575A..55A,2016ApJ...819..121P}.  At kiloparsec scale, it presents a compact structure with no diffuse emission \citep{2018A&A...614A..87B}, while on parsec scale it shows a core-jet structure with superluminal motion and apparent velocity up to $\sim8c$ (e.g., \citealt{2016AJ....152...12L}).  Regarding its host galaxy, no significant structure was detected with the {\it Hubble Space Telescope} ({\it HST}\,) \citep{1993ApJ...402...69M} in an observation made before the telescope's spherical aberration was corrected.  Our new data are the first high-resolution observations of this source.

\section{Observations}\label{sec:observations}

\subsection{Imaging}

We observed SBS~$0846+513$ with the LUCI1 camera for two nights on 2016 December 14-15, as part of the commissioning run for ARGOS.  LUCI1 and LUCI2 are twin instruments designed for both spectroscopy and imaging. In the imaging mode, they provide a 4$\times$4 arcmin field of view.

On December 14, we observed with the infrared {\it Ks} and {\it J} filters, but the seeing was bad and the images unusable.  On December 15, we observed with the {\it J} filter only.  On this night, the seeing and image quality were good, and we obtained a total exposure time of 31.7 minutes for the science observations, with individual exposures of 10 s each.   The image scale is 0.118 arcsec per pixel.  Building up a composite Point Spread Function (PSF) from stars in the field, we obtain a PSF Full Width at Half-Maximum (FWHM) of 0.308 arcsec.  The image reduction and PSF construction are described in \S\ref{sec:imageprocessing}

\begin{figure}
	\includegraphics[width=\columnwidth]{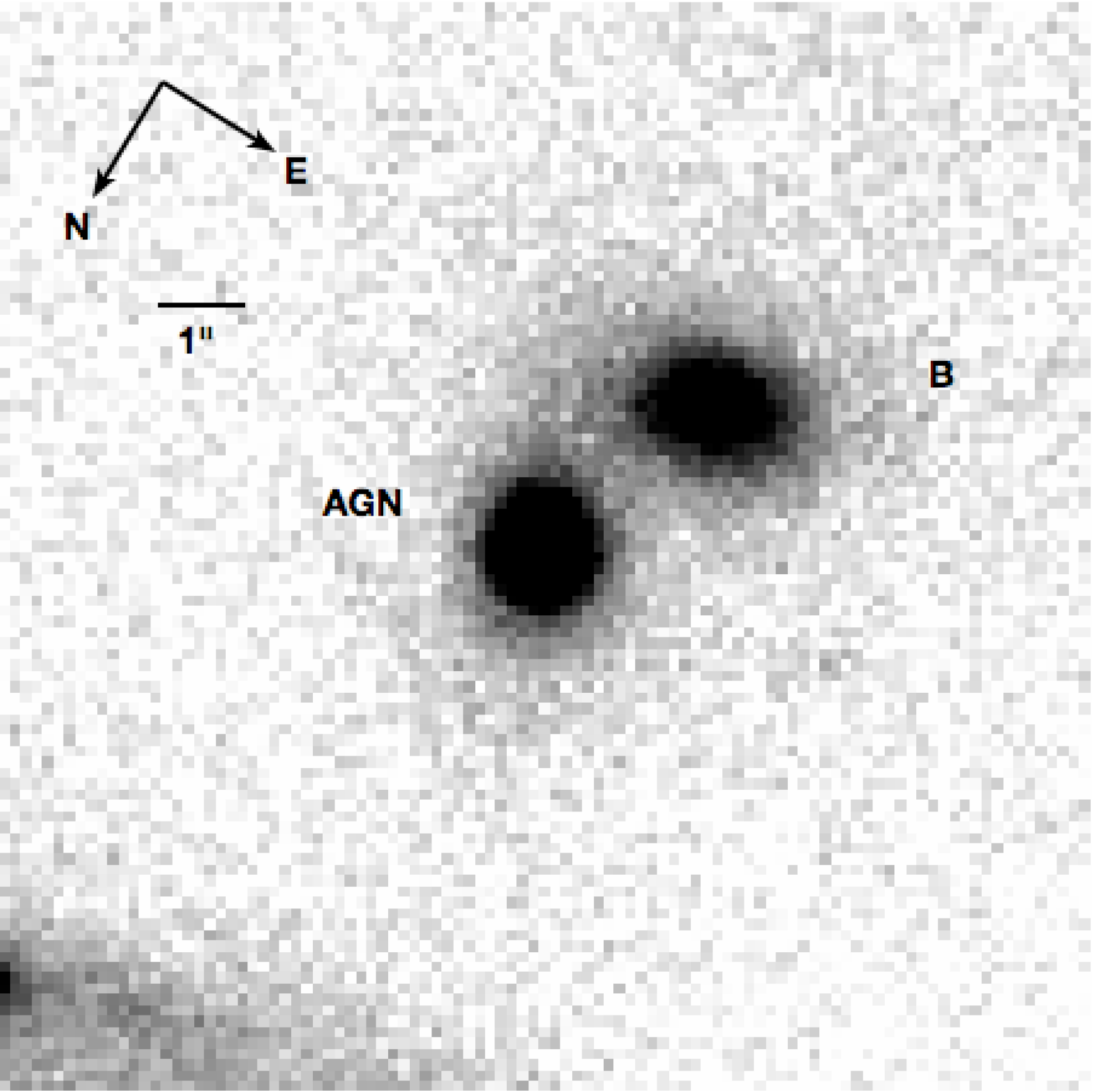}
	\caption{$J$-band view of the field surrounding SBS~$0846+513$.  SBS~$0846+513$ is marked `AGN,' and the projected companion `B.'  At bottom, an arm from a large, foreground spiral can be seen.  The scale bar spans 1 arcsec.}
	\label{fig:field-annotated}
\end{figure}

The target region is shown in Fig.~\ref{fig:field-annotated}.  The NLS1, SBS~$0846+513$, is marked `AGN.'  Partially overlapping the AGN is a second galaxy, labelled `B' in the image, that has a projected distance of 2.45 arcsec centre-to-centre.  

\subsection{Spectroscopy and Redshift}\label{sec:spectroscopy}

Because Galaxy B is projected to partially overlap the AGN, it raises the question of whether or not B is a physical companion, and possibly interacting.  We addressed this spectroscopically, observing SBS~$0846+513$ with the Alhambra Faint Object Spectrograph and Camera (ALFOSC) on the Nordic Optical Telescope (NOT) on May 9, 2018. The spectrum was obtained with a 1 arcsec slit oriented at position angle $160^{\circ}$, to fit both SBS~$0846+513$ and Galaxy B in the slit at the same time. The spectrum interval covered was between 5650 and 10150 \AA, with a resolution R$\sim$770 and a total exposure time of 1800 s. After the standard bias and flat-field correction, the wavelength calibration was performed using a ThAr lamp. We flux calibrated the spectrum using as reference the standard star HD 93521.

\begin{figure}
	\includegraphics[width=\columnwidth]{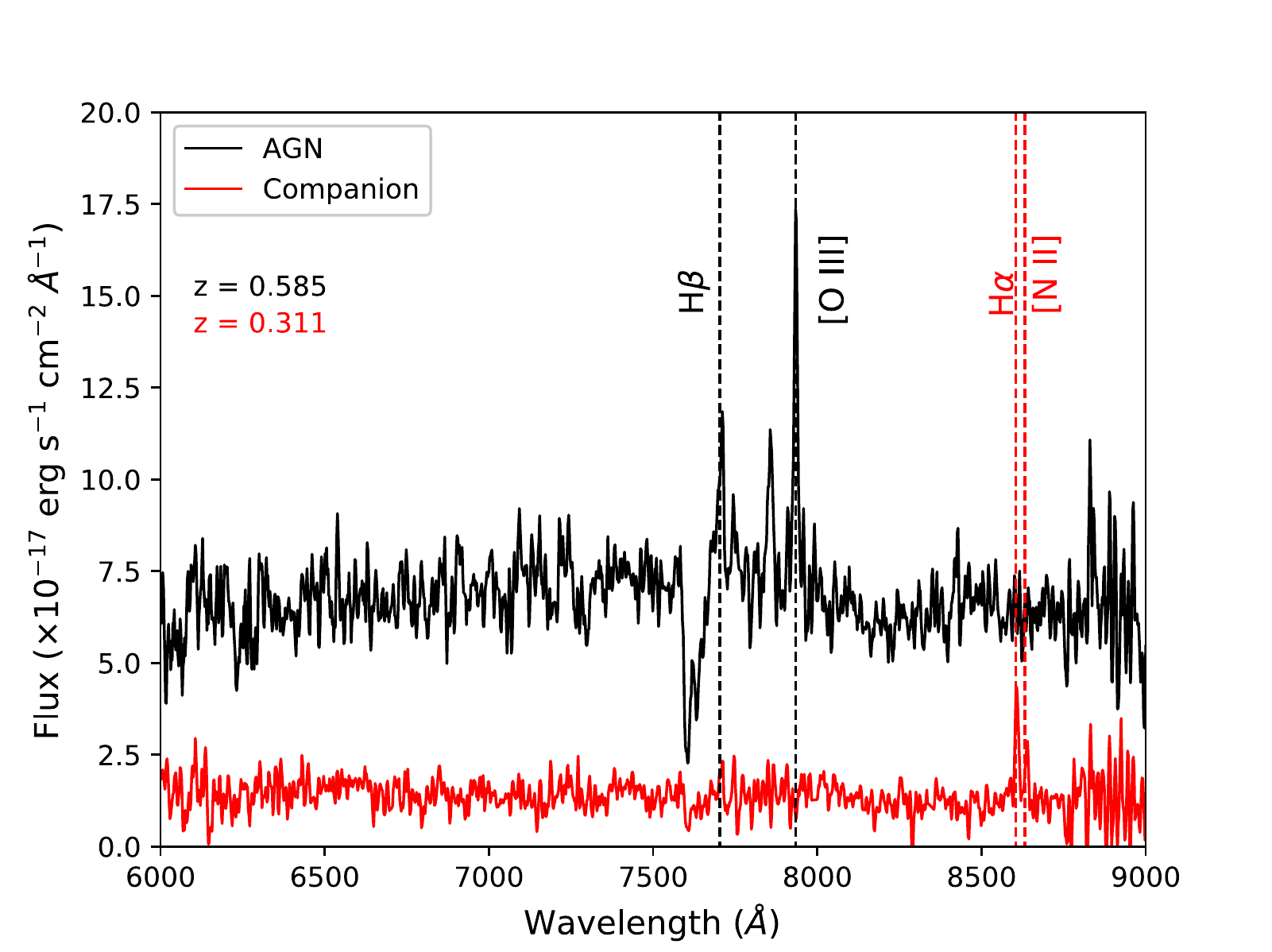}
	\caption{NOT spectra of the AGN (upper line, black) and the apparent companion (lower line, red).  Identified emission lines are marked with dashed, vertical lines: H~$\beta$ and [O\,\textsc{iii}] in the AGN, H$\alpha$ and [N\,\textsc{ii}] in the apparent companion.  We measure a redshift of $z=0.585$ for the AGN and $z=0.311$ for Galaxy B.}
	\label{fig:spec}
\end{figure}

The spectrum of the two sources is shown in Fig.~\ref{fig:spec}. We clearly identify the AGN lines at $z = 0.585$, as expected. Both H~$\beta$ and [O\,\textsc{iii}]$\lambda\lambda$ 4959, 5007 are clearly visible, next to the telluric absorption by the atmosphere. In the putative companion, instead, neither line is visible at the same wavelength. Instead, we detected at 8604 and 8631 \AA, two narrow emission lines compatible with redshifted H$\alpha$ and [N\,\textsc{ii}]$\lambda$6584 at $z = 0.311$.  In principle, the H~$\beta$ region and the [O\,\textsc{iii}] 5007 \AA{} line would also be visible in Galaxy B, but they are not detected in the spectrum, possibly suggesting the presence of internal absorption.  Regardless of the reason these are not detected, Galaxy B is definitely not at the same redshift as SBS~$0846+513$.

\section{Analysis}\label{sec:analysis}

\subsection{Image Processing}\label{sec:imageprocessing}

The data reduction process for producing the science image follows the steps described by \citet{2019MNRAS.484.3356G} in their ARGOS commissioning paper.  After a basic reduction, correcting for non-linearity and persistence effects, bad pixel masking and flat fielding, sky frames were created from exposures spaced about 10 minutes apart in order to negate effects by the rapidly-varying Near-Infrared (NIR) atmosphere.  The best 164 exposures were used in this case.  Bright sources, such as stars and galaxies (with pixel values $\sim \!4\, \sigma$ above the background) were masked prior to combination.  Several sky frames were used to subtract the sky in a group of exposures spaced within about 15 minutes.

Due in part to the extended spiral galaxy nearby and its effect on the flatfielding, the sky in source-free areas varies somewhat in the region around the AGN.  At 15 arcsec east of the AGN, it has a mean of 1.47 Analog-to-Digital Units (ADU), 16 arcsec west of the AGN, the mean is 2.24 ADU, and 22 arcsec south of the AGN, the mean is 3.03 ADU.  The large galaxy is to the north of the AGN, so no measurement is made there.  The statistics for these regions is listed in Table~\ref{tab:sky-stats}.  To minimize these effects in the analysis, we restrict the model fit (\S\ref{sec:hostmodel}) to a region close to the AGN.

\begin{table}
\caption{Statistics of sky and residuals.  
Columns--
(1) Region or model residual. South, East, and West refer to source-free regions near the AGN.  1C and 2C residuals have the total model subtracted from the image and use the fitting region centreed on the AGN.
(2) Mean value.
(3) Minimum value.
(4) Maximum value.
(5) Standard deviation.
(6) Distance of region from the AGN.
Note $a$--Modeled sky level for the 1C model is 1.24 ADU.
Note $b$--Modeled sky level for the 2C model is 1.73 ADU.
}
\label{tab:sky-stats}
\begin{tabular}{|c|ccccc|}\hline
(1)  		&(2)   	&(3)    &(4)    &(5)	&(6)\\
Region		&Mean	&Minimum&Maximum&$\sigma$&Distance\\ 
			&(ADU)	&(ADU)	&(ADU)	&(ADU)	&(arcsec)\\
\hline
East 		&1.47	&-18.41	&21.81	&4.52	&15\\
West		&2.24	&-20.32 &16.1	&4.41	&16\\
South		&3.03	&-17.02	&17.25	&4.35	&22\\
1C residual	&$0.00^{\,a}$   &-83.49 &101.12 &5.79	&0\\
2C residual	&$0.00^{\,b}$   &-65.23 &97.17  &5.58	&0\\
\hline
\end{tabular}
\end{table}

\subsection{Photometric Calibration}

We calibrated the {\it J}-band magnitude system using the 2-Micron All Sky Survey (2MASS) Extended Source Catalog and Point Source Catalog.  Instrumental magnitudes for sources from our image were extracted with Source Extractor and compared to the 2MASS catalogs.  Excluding the AGN (which is variable), Galaxy B (which is merged with the AGN in 2MASS), and one bright star that is saturated, nine sources in this field are found in the Point Source Catalog (PSC).  Two of these are spiral galaxies and are also listed in the Extended Source Catalog (XSC).  We adopt the XSC $J$ magnitudes for these two and the PSC $J$ magnitudes for the remaining seven.  With these sources, we obtain a magnitude calibration with a standard deviation of $\sigma=0.26$ mag.

\subsection{Host + Nucleus Modeling}\label{sec:hostmodel}

Because this active galaxy has a bright nucleus, any study of its morphology requires a fit to both the nucleus--modeled with the point spread function (PSF)--and the host galaxy.  We use \textsc{galfit} \citep{2011ascl.soft04010P} to perform a 2-dimensional fit to the image, simultaneously modeling the nucleus, the host galaxy, and Galaxy B, with the sky fitted as a constant value.  Because of the spatial variations in the sky described in \S\ref{sec:imageprocessing}, we perform the fit over a region spanning $101\times101$ pixels (11.9$\times$11.9 arcsec), centred on the AGN.  This is wide enough to cover the visible extent of the AGN host and Galaxy B but narrow enough so that the sky variation within the region is small.  The large, foreground spiral seen in the north corner of Fig.~\ref{fig:field-annotated} and the region of brighter sky to the south of Galaxy B (which may include a low-surface-brightness source) are both masked out of the fit, as are smaller, dimmer clumps.  The PSF model is constructed as a composite of 16 isolated stars from across the field of view and has a Full Width at Half Maximum of 2.6 arcsec.  It spans $19\times19$ pixels (2.2$\times$2.2 arcsec) and is put into \textsc{galfit} with normal sampling.

\begin{figure}
\includegraphics[width=\columnwidth]{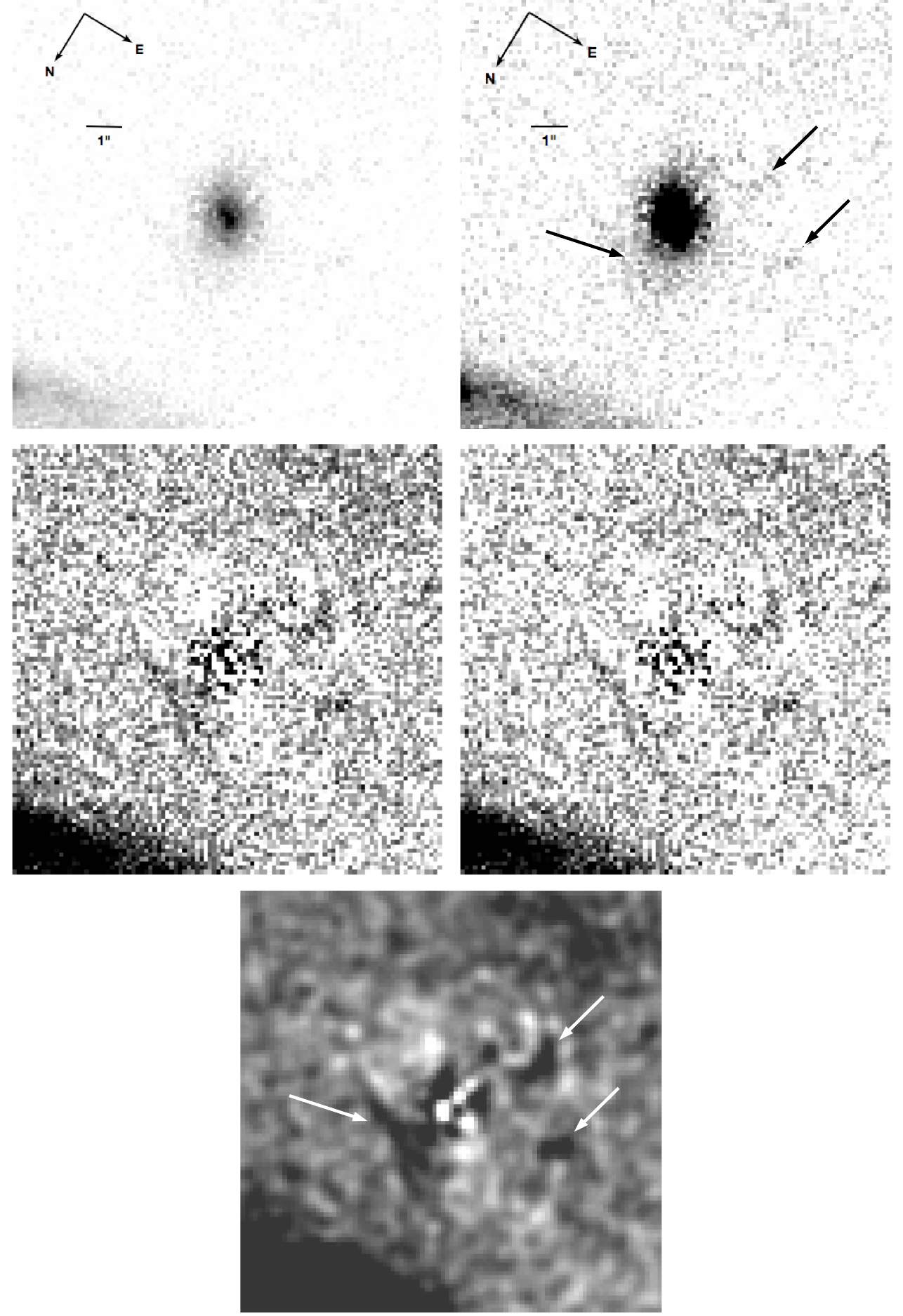}
\caption{View of the AGN host galaxy and residuals.  The field of view is the same as in Fig.~\ref{fig:field-annotated}.  {\it Top row:} The PSF and Galaxy B models have been subtracted, showing the AGN host galaxy.  {\it Top left:} Low-contrast image, showing the inner features of the host.  The noise 1 arcsec from the host centre is due to scaling up the faint edges of the stellar PSF model.  {\it Top right:} High-contrast image, emphasizing the outer envelope.  Three clumps of material are indicated by the arrows.  The lower two are just visible in Fig.~\ref{fig:field-annotated} but are clearer here.  The clump to the southeast of the AGN is hidden by Galaxy B in the original image.  {\it Middle row:} Residuals made by subtracting the total model.  An inverted color scheme (blacker is brighter) is used, in keeping with the images at top.  {\it Middle left:} 1C model residual. {\it Middle right:} 2C model residual.  The higher noise at the centre, extending to a radius of 1 arcsec, is due to the PSF model.  The apparent structures noted in the top row stand out above the variations in the sky and residual.
{\it Bottom:} 1C model residual with 3-pixel Gaussian smoothing.  The clumps of material noted before (marked with arrows) are more visible.
}
\label{fig:host-residual}
\end{figure}

To determine the morphology, two models of the host galaxy are tried.  Model 1C is a one-component fit to the AGN host using a S\'{e}rsic model, while model 2C performs a two-component fit to the host, using a S\'{e}rsic bulge and an exponential disc.  The S\'{e}rsic index, $n$, is a free parameter in both cases.  Since we are not interested in the details of its morphology, Galaxy B is modeled with only single S\'{e}rsic component (with the index as a free parameter) in both cases, which leaves acceptably low residuals.  Including separate bulge and disc components results in unreasonable fits.  The sky is modeled as a constant value; allowing it to `tilt' across the field does not improve the fit.  The host galaxy is shown in the PSF-subtracted residual image in Fig.~\ref{fig:host-residual}, along with the total residuals, formed by subtracting the complete model from the image.  The fitting results are summarized in Table~\ref{tab:fit-agn} for the AGN and Table~\ref{tab:fit-b} for Galaxy B, and the statistics of the residuals are listed in Table~\ref{tab:sky-stats}.

Absolute magnitudes and physical scales for both galaxies are presented in Table~\ref{tab:derived}, using a cosmology of $H_0=71$, $\Omega_\mathrm{M}=0.27$, and $\Omega_\mathrm{vac}=0.73$.  At the AGN's redshift, the scale is 6.59 kpc/arcsec, while for Galaxy B, it is 4.53 kpc/arcsec.  To calculate the absolute magnitudes, we adopt $K$-corrections for the AGN nucleus from \citet{2006ApJ...640..579G} and for the AGN host and Galaxy B from \citet{2001MNRAS.326..745M}.  For the purposes of the $K$-correction, we assume the AGN host is an E-type galaxy for the 1C model and an Sa-type galaxy for the 2C model, while Galaxy B is treated as an Sa type.

\begin{figure}
\includegraphics[width=\columnwidth]{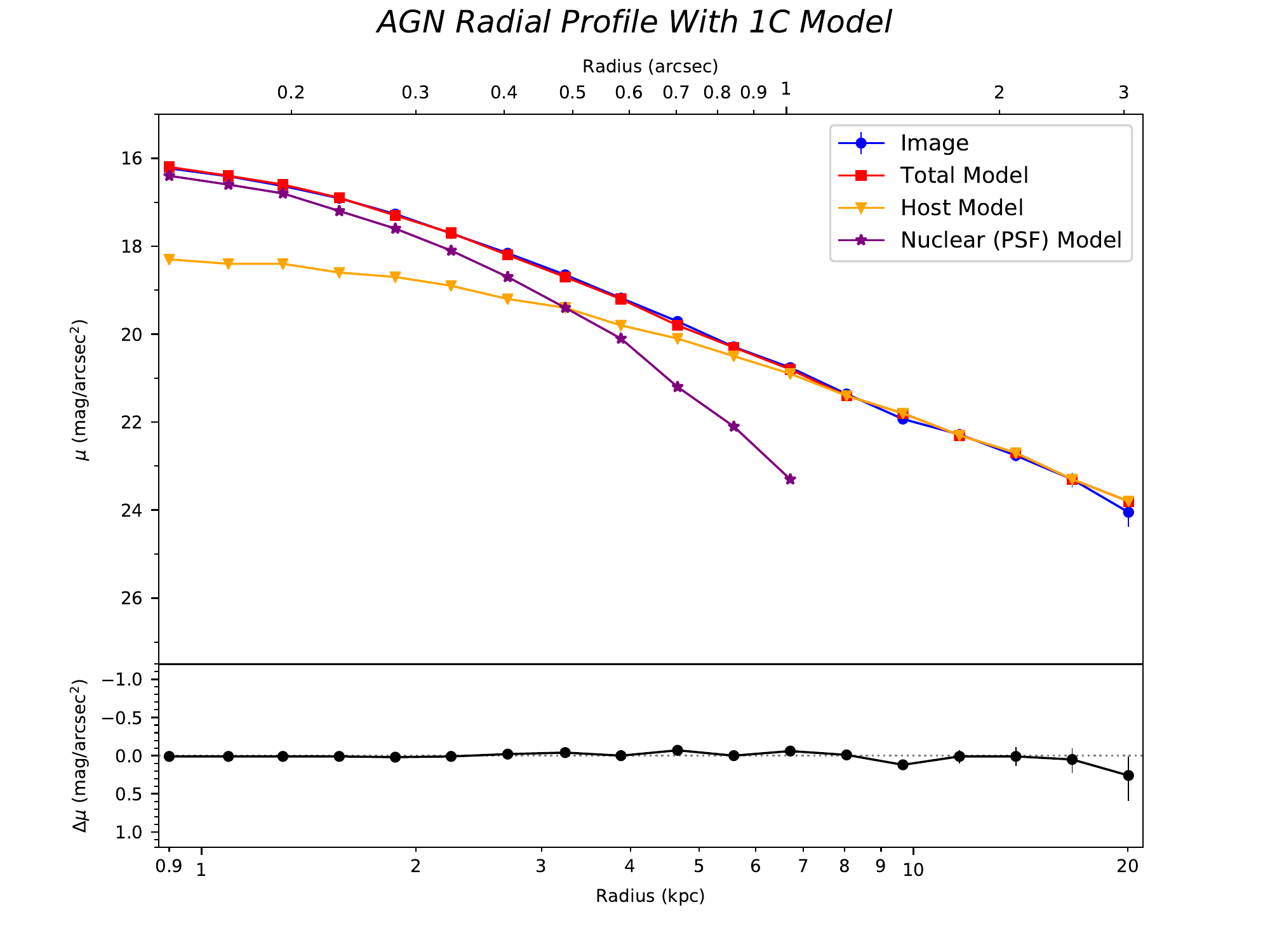}
\caption{Radial surface brightness ($\mu$) plot of the AGN image and components for the 1C (S\'{e}rsic) model, using circular annuli.  The upper plot shows the profiles of the image (with the sky and the Galaxy B model removed), the total model, and the model's individual components.  The lower plot shows the difference in surface brightness between the image and the total model ($\Delta\mu = \mu_\mathrm{image} - \mu_\mathrm{total\, model}$).  The horizontal axis is marked in both arcseconds (top scale) and projected kpc for the redshift of the AGN (bottom scale) for both plots.  The vertical range is scaled to match Fig.~\ref{fig:radial-2c} for ease of comparison.  Error bars are given for the image and $\Delta \mu$.  
The single S\'{e}rsic model provides a good match to the profile out to about 3 arcsec (20 kpc).  Beyond this, the image profile is lost in the sky variation.
}
\label{fig:radial-1c}
\end{figure}

\begin{figure}
\includegraphics[width=\columnwidth]{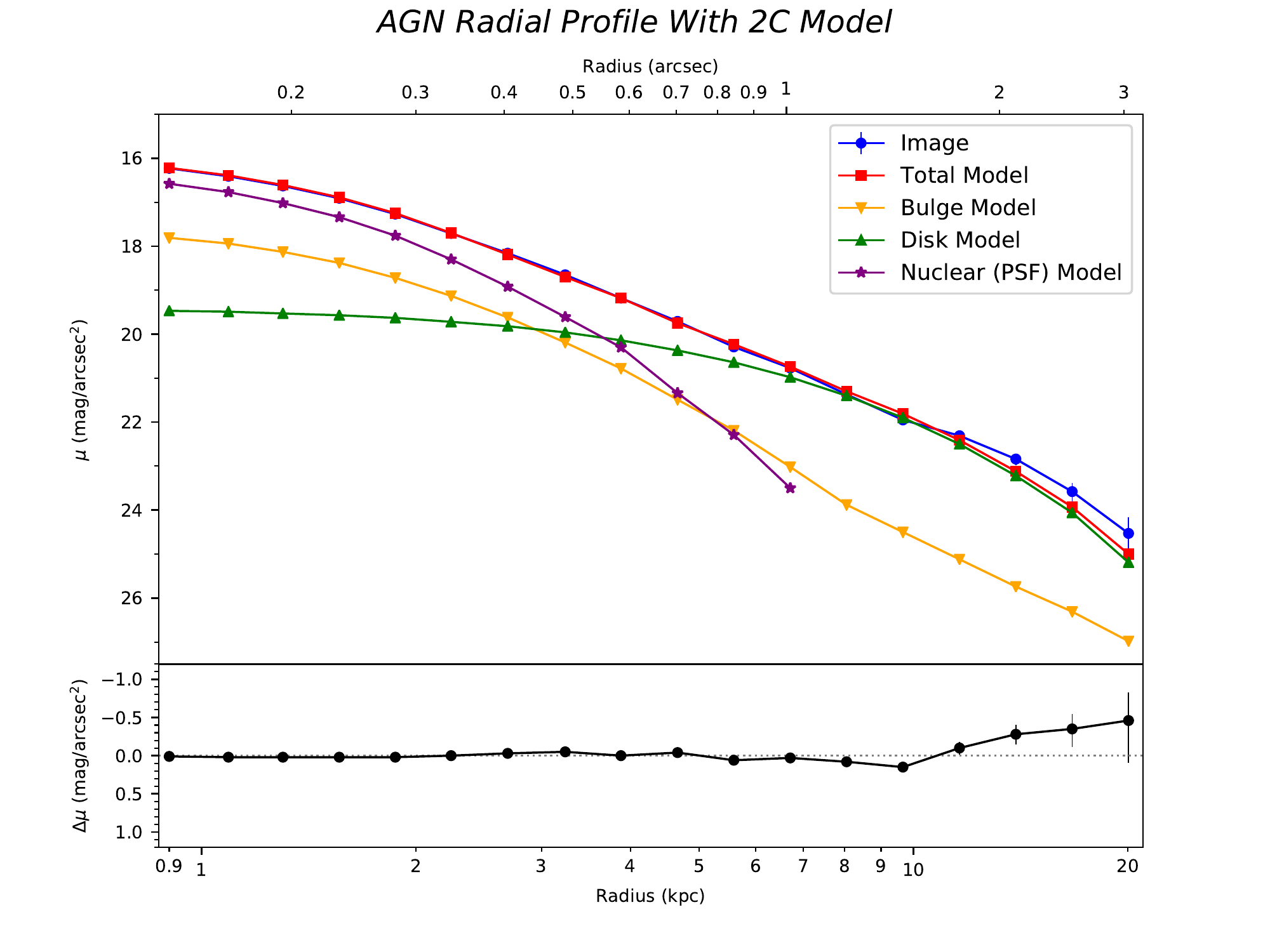}
\caption{Radial surface brightness ($\mu$) plot of the AGN image and components for the 2C (S\'{e}rsic bulge + exponential disc) model.  The plot layout and scale are the same as in Fig.~\ref{fig:radial-1c}.  The S\'{e}rsic + exponential disc model is a good match to the profile out to about 1.8 arcsec (12 kpc).  From there to about 3 arcsec (20 kpc), there is excess light from the filament-like structure to the north and imperfectly-subtracted light in Galaxy B, both shown in the residual image in Fig.~\ref{fig:host-residual}.  Though the 2C model's radial profile diverges from the image in the outer radii, it provides a better overall fit in two dimensions than the 1C model does.
}
\label{fig:radial-2c}
\end{figure}

In addition to the two-dimensional modeling, we examine the AGN's radial profile and isophote structure.  Fig.~\ref{fig:radial-1c} and Fig.~\ref{fig:radial-2c} show the radial surface brightness ($\mu$) profiles of the AGN for the 1C and 2C models, respectively, measured using \textsc{iraf}'s {\tt ellipse} task with circular annuli.  Plots are included of the image (with the sky and the Galaxy B model subtracted), the total model, and the separate components of that model--the host model (1C: S\'{e}rsic only; 2C: S\'{e}rsic bulge and exponential disc) and the nuclear model (stellar PSF).  The profile is taken out to an angular distance of 3 arcsec from the centre of the nucleus, corresponding to a projected distance of 20 kpc at the redshift of the source.  The bottom portion of the figure shows the difference between the image and the total model for each isophote: $\Delta\mu = \mu_\mathrm{image} - \mu_\mathrm{total\, model}$.

\begin{figure}
\includegraphics[width=\columnwidth]{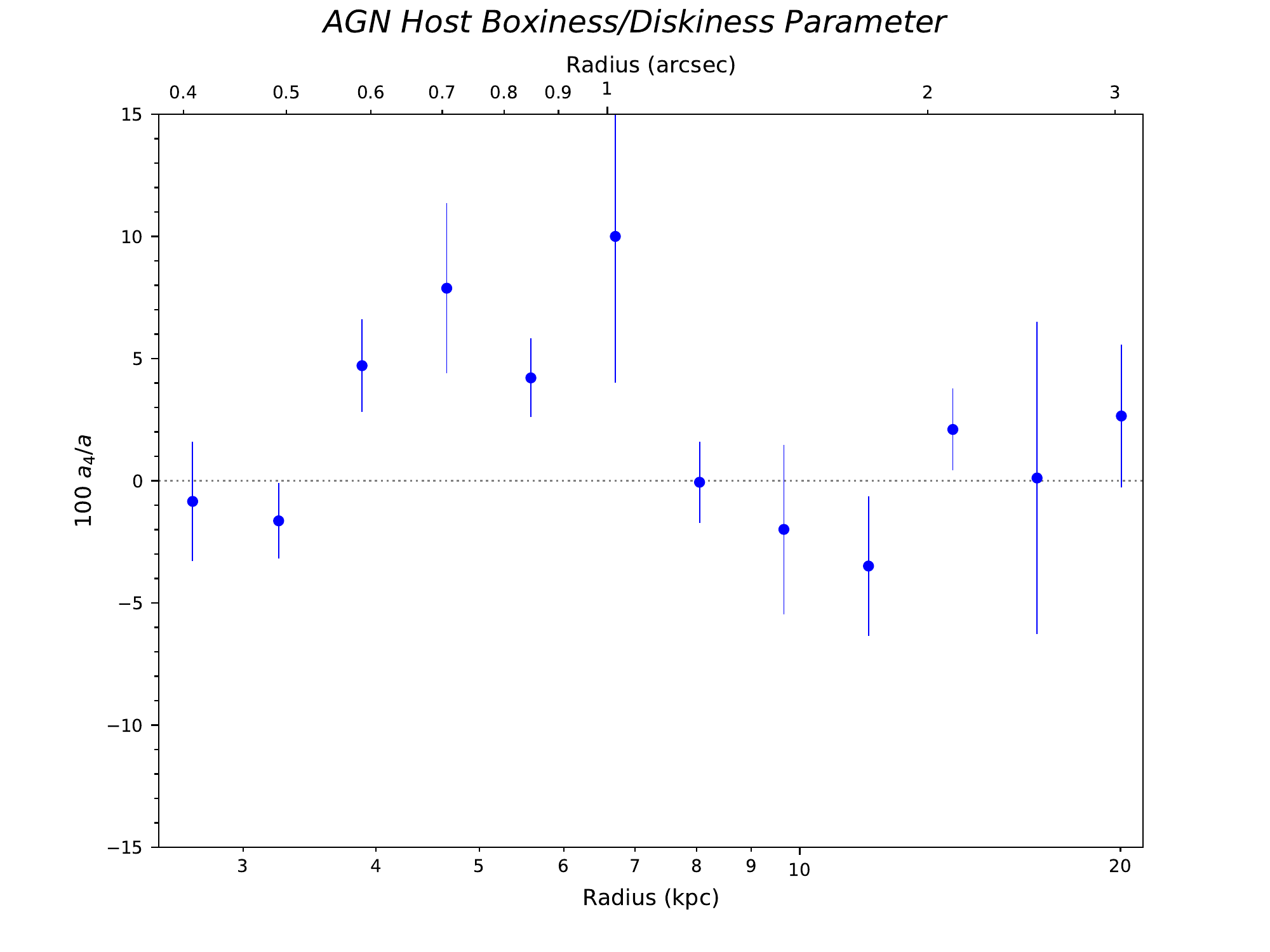}
\caption{Isophote boxiness/disciness parameter as a function of semi-major axis.  Negative values of $a_4<0$ indicate boxiness, while positive values indicate disciness.  Using the criteria of \citet{1989A&A...217...35B}, we can assign a characteristic value of $a_4/a\sim 0.05 - 0.08$, which is discy.
}
\label{fig:iso_a4}
\end{figure}

In Fig.~\ref{fig:iso_a4}, \textsc{iraf}'s {\tt ellipse} task is also used to measure the Fourier $a_4/a$ parameter of the isophotes of the PSF-subtracted host.  Negative values of $a_4/a$ indicate boxiness in a given isophote, a more rectangular deviation from an ellipse, while positive values indicate discy, lemon-shaped isophotes.  An overall discy trend to the isophotes is associated with the presence of galactic discs.  The convolution of the bright centre of the galaxy with the PSF  biases the shape of the innermost isophotes, so we do not plot isophotes closer than 0.4 arcsec (2.6 kpc) from the centre.  The vertical axis is scaled by a factor of $100\times$, and the dashed horizontal line at $a_4/a=0$ separates the positive (discy) values from negative (boxy) values.

\begin{table*}
\caption{Summary of AGN model fits.  Columns--(1) Model type: 1C uses a one-component S\'{e}rsic fit to the AGN host, listed under `Bulge,' because of its high S\'{e}rsic index. Model 2C uses two AGN host components--a S\'{e}rsic bulge and an exponential disc.  
(2) Reduced $\chi^2$ of the fit.
(3) Nuclear $J$ magnitude.  
(4)--(11) $J$ magnitude ($m$), radius ($r$), S\'{e}rsic index ($n$), and ellipticity ($\epsilon$) for host bulge and disc.  
1-$\sigma$ uncertainties are listed in brackets.  
Parameters that are held fixed are listed in parentheses.  Subscripts $p$, $b$, and $d$ refer to nucleus (PSF), bulge, and disc, respectively.}
\label{tab:fit-agn}
\begin{tabular}{|cc|c|cccc|cccc|}\hline
(1)  &(2)         &(3)           &(4)           &(5)           &(6)           &(7)                  &(8)           &(9)           &(10)          &(11) \\
Model&$\chi^2/\nu$&Nucleus (PSF)       &\multicolumn{4}{c|}{Bulge}	                                      &\multicolumn{4}{c|}{disc} \\
     &            &$m_\mathrm{p}$&$m_\mathrm{b}$&$r_\mathrm{b}$&$n_\mathrm{b}$&$\epsilon_\mathrm{b}$&$m_\mathrm{d}$&$r_\mathrm{d}$&$n_\mathrm{d}$&$\epsilon_\mathrm{d}$ \\
	 &            &(mag)	     &(mag) 	    &($\arcsec$)   &              &                     &(mag) 	       &($\arcsec$)   &              & \\ \hline
1C	 &1.2         &20.3	[$<0.1$] &20.3 [$<0.1$] &1.1 [0.1]     &3.4           &0.41                 &--           &--           &--           &--  \\
2C	 &1.1         &20.4 [$<0.1$] &21.3 [$<0.1$] &0.1 [$<0.1$]  &7.3           &0.86                 &20.9 [$<0.1$] &0.6 [$<0.1$]  &(1)           &0.30 \\
\hline
\end{tabular}
\end{table*}

\begin{table}
\caption{Summary of model fits for Galaxy B.  
Columns--(1) Model type (same as above).  
(2) Apparent $J$ magnitude.  
(3) Half-light radius.  
(4) S\'{e}rsic index.  
(5) Ellipticity. 
1-$\sigma$ uncertainties are listed in brackets. 
}
\label{tab:fit-b}
\begin{tabular}{|c|cccc|}\hline
(1) 	& (2) 		     & (3) 			  & (4) 		   & (5)  					  \\
Model	& $m_\mathrm{B}$ & $r_\mathrm{B}$ & $n_\mathrm{B}$ & $\epsilon_\mathrm{B}$	  \\
		& (mag)	         & ($\arcsec$)	  & 	           & 			              \\ \hline
1C		& 20.1	[$<0.1$] & 1.2 [$<0.1$]   & 2.3			   & 0.33					  \\
2C		& 20.1  [$<0.1$] & 1.1 [$<0.1$]   &	2.2  		   & 0.34					  \\
\hline
\end{tabular}
\end{table}

\begin{table*}
\caption{Derived quantities for the AGN and Galaxy B.
Columns--(1) Model (same as above).  
(2) $J$-band absolute magnitude of the AGN nucleus.  
(3)--(6) $J$-band absolute magnitudes ($M$), and radii ($r$), for the AGN bulge and disc.  
(7)--(8) Bulge-to-Total ratio and total absolute magnitude for the AGN host.  
(9)-(10) $J$-band absolute magnitude and radius for galaxy B.
1-$\sigma$ uncertainties are listed in brackets.
Subscripts are the same as above.
}
\label{tab:derived}
\begin{tabular}{|c|c|cc|cc|cc|cc|}\hline
(1)  &(2)             &(3)           &(4)           &(5)           &(6)           &(7)  &(8)              &(9)           &(10) \\
Model&Nucleus 	      &\multicolumn{2}{c|}{Bulge}   &\multicolumn{2}{c|}{disc}    &     & &\multicolumn{2}{c|}{Galaxy B}  \\
     &$M_\mathrm{n}$  &$M_\mathrm{b}$&$r_\mathrm{b}$&$M_\mathrm{d}$&$r_\mathrm{d}$&$B/T$&$M_\mathrm{host}$&$M_\mathrm{B}$&$r_\mathrm{B}$ \\
     &(mag)           &(mag)         &(kpc)         &(mag)         &(kpc)         &     &(mag)            &(mag)         & (kpc)    \\ \hline
1C   &-22.4           &-22.7         &7.5 [0.5]     &--           &--           &1.00 &-22.7            &-21.1         &5.2 [0.2] \\
2C   &-22.3           &-21.6         &0.7 [0.1]     &-22.0         &4.0 [0.1]     &0.41 &-22.6            &-21.1         &5.1 [0.2] \\
\hline
\end{tabular}
\end{table*}

\subsection{Black Hole Mass}\label{sec:bhmass}

We use three methods to determine the central black hole mass--the correlation with the host bulge, and two correlations with the S\'{e}rsic index.  Detailed results are shown in Table~\ref{tab:bhmass}.  The relationship between host spheroidal luminosity and black hole mass is given by \citet{2003ApJ...589L..21M} as
\begin{equation}
	\log\frac{\text{M}_{\mathrm{BH}}}{\text{M}_\odot} = a + b X \mathrm{ ,}
	\end{equation}
where for the $J$-band, $a = 8.26 \pm 0.07$, $b = 1.14 \pm 0.12$, $X = \log L_{J \mathrm{, bulge}} - 10.7$, $\text{M}_{\mathrm{BH}}$ represents the black hole mass, $\text{M}_\odot$ is the Sun's mass, and $L_{J \mathrm{, bulge}}$ is the $J$-band spheroidal luminosity.  In terms of spheroidal magnitude, $M_{J, \mathrm{bulge}}$, this becomes
\begin{equation}
	\log\frac{\text{M}_{\mathrm{BH}}}{\text{M}_\odot} = 
	0.46 ( M_{J,\odot} - M_{J, \mathrm{bulge}} ) - 3.94 \,\mathrm{ ,}
	\end{equation}
Using $M_{J,\odot} = 3.67$ for the Sun's magnitude, we obtain a black hole mass of $1.55\times10^8 \,\text{M}_\odot$ for the 1C model and $4.97\times10^7 \, \text{M}_\odot$ for the 2C model.

The radial profile method of \citet{2007ApJ...655...77G} correlates the S\'{e}rsic index, $n$, with the black hole mass.  They fit both linear (their Equ.~4) and quadratic (their Equ.~7) relationships to the data.  The linear fit gives
\begin{equation}
	\log \frac{\text{M}_{\mathrm{BH}}}{\text{M}_\odot} = 
	(2.68 \pm 0.40) \log \frac{n}{3} + (7.82 \pm 0.07) \, \mathrm{ ,}
	\end{equation}  
while the quadratic fit is
\begin{multline}
	\log \frac{\text{M}_{\mathrm{BH}}}{\text{M}_\odot} = 
	(7.98 \pm 0.09) + (3.70 \pm 0.46) \log\frac{n}{3} - \\
	(3.10 \pm 0.84) \left( \log\frac{n}{3} \right)^2  \mathrm{ .}
	\end{multline}
As shown in Table~\ref{tab:bhmass}, model 1C leads to a mass of $9.09\times10^7 \, \text{M}_\odot$ for the linear fit or $1.46\times10^8 \, \text{M}_\odot$ for the quadratic fit.  Since the two-component model bulge is smaller than the PSF width, we do not calculate profile-derived masses for model 2C.

\begin{table}
\caption{Central black hole mass of the AGN, calculated by different methods: `Bulge luminosity' is the black hole mass/bulge luminosity relation; `S\'{e}rsic linear' is the linear fit to the black hole mass/S\'{e}rsic index relation; `S\'{e}rsic quadratic' is the quadratic fit to the latter.  Models are the same as above.  Mass uncertainties (listed in brackets) account for uncertainties in the proxy and statistical uncertainties in the input parameter, but they cannot account for systematic errors in the input parameter.  }
\label{tab:bhmass}
\begin{tabular}{|c|cc|}\hline
Method               & Model & $\log \left( \text{M}_\mathrm{BH}/\text{M}_\odot \right)$  \\ \hline
Bulge luminosity     & 1C    & 8.19 [0.08] \\
Bulge luminosity     & 2C    & 7.70 [0.10] \\
S\'{e}rsic linear    & 1C    & 7.96 [0.12] \\
S\'{e}rsic quadratic & 1C    & 8.16 [0.13] \\ \hline
\end{tabular}
\end{table}

\section{Discussion}\label{sec:discussion}

\subsection{Host Galaxy}\label{sec:host}

In interpreting the relationship of these gamma-ray emitters to the broader population of NLS1, it is important to compare their morphologies.  But very few of the gamma-ray emitters' hosts have been resolved, because the fraction of NLS1 known to emit gamma-rays peaks at $z\sim0.5-0.7$, where their hosts become harder to distinguish, and high-resolution observations such as these are necessary.    Narrow-Line Seyfert 1s at low redshift have been found to reside primarily in spiral galaxies, similar to the overall population of Seyfert 1s.  \citet{2003AJ....126.1690C} and \citet{2006AJ....132..321D} have only one clear E type in their NLS1 samples. \citet{2017MNRAS.467.3712O} has one classified as SB0.

How does SBS~$0846+513$'s morphology compare to lower-redshift NLS1?
As shown in Table~\ref{tab:fit-agn}, the one-component (1C) and two-component (2C) models of the AGN have similar goodness of fit, with a small improvement in the 2C model.  The one-component model finds the AGN host galaxy to be roughly of the de Vaucouleurs profile, with an overall S\'{e}rsic index of 3.4.  The two-component model uses an exponential disc and finds the bulge to have a steeper S\'{e}rsic index of 7.3, but the bulge is smaller than the PSF width, so we cannot do much interpretation of it.  The bulge-to-total ratio of 0.41 (and to the extent we can interpret it, the bulge S\'{e}rsic index) is consistent with an Sab Hubble type \citep{2009ApJ...696..411W}.  Using the 6dF luminosity function of \citet{2006MNRAS.369...25J} and a Hubble constant of $H_0=71$ km/s/Mpc, we obtain $M^*_J=-23.6$, while the host galaxy of SBS~$0846+513$ has a total magnitude of -22.7 (1C model) or -22.6 (2C)\footnote{The small discrepancy comes from the different $K$-corrections for the morphologies.}.  This puts the host at 1 mag fainter than $M^*_J$.

The radial profile in Fig.~\ref{fig:radial-1c} shows a good fit of the 1C model to the image out to 3 arcsec (20 kpc).  Beyond this point, the variations in the background produce some regions of negative values in the sky-subtracted image, which we can't display in magnitude form.  The 2C model (Fig.~\ref{fig:radial-2c}) gives a close fit over most of the isophotes but shows some divergence at the outermost radii, where the model is too faint.  Still, the 2C model has a slightly better reduced $\chi^2$ than the 1C model in the two-dimensional fit.  

We next examine the radial variation of the isophotes' $a_4/a$ parameter (Fig.~\ref{fig:iso_a4}), which is associated with boxy (negative) or discy (positive) isophotes.  We disregard the central 0.3 arcsec, where the isophote shapes are likely to be dominated by convolution with the PSF.  We see that $a_4/a$ has a roughly-peaked distribution, rising from $\sim0$ close to the centre, to large positive values in the $4 < r < 7$ kpc region, beyond which it declines irregularly.  Though the graph is not smooth, under the criteria of \citet{1989A&A...217...35B}, we can assign a characteristic value of $a_4/a\sim0.08$, and this strengthens the evidence for a disc.

In Fig.~\ref{fig:host-residual}, we show residuals created by subtracting the PSF and Galaxy B models (top row) and the total model (middle and bottom rows) from the image.  In the top row, we can see more clearly some clumps of material, marked by arrows.  The lower two are just visible in Fig.~\ref{fig:field-annotated} but are clearer in the residual, while the third (southeast of the AGN) is normally covered by Galaxy B until that model is subtracted.  The extended features to the northwest and southeast of the AGN could be spiral arms.  They're on opposite sides, at about the same distance from the nucleus, but their shapes aren't symmetrical and could show a disturbance.  These three features are only of low significance ($\lesssim1\sigma$ above sky level), and it would be interesting to see them in a deeper image.

Another interpretation of the hook-shaped pattern southeast of the AGN could be as a feature of Galaxy B.  The outer isophotes of this galaxy have a slightly asymmetrical appearance to the eye, extending a bit farther to the west than the east.  Our attempts to account for this in \textsc{galfit} (such as adding Fourier components or a non-concentric bulge and disc) either failed to improve the fit or kept it from converging.  

The residual statistics for the unmasked regions are listed in the bottom two rows of Table~\ref{tab:sky-stats}.  There is no systematic offset from the sky level, though there is a higher standard deviation compared to the neighboring source-free regions.  Some of this comes from noise in the PSF model for the nucleus, which is scaled in brightness during the fit.  The noisy disc at the centre of the fitting region in Fig.~\ref{fig:host-residual} (middle row) results from the scaled-up PSF noise and extends out to a radius of 1 arcsec.  The 2C model leaves a residual with a slightly lower standard deviation than the 1C model does, but then its PSF (which is a source of noise) is fainter by 0.1 mag.    The brighter region to the south of Galaxy B, an area masked from the model fit, includes some low-surface-brightness sources that are just outside the cropped field.

So in its morphology, SBS~$0846+513$ does not appear obviously different from the common spiral galaxies of low-redshift NLS1.  The bulge + disc model fits it slightly better than the spheroidal model, the isophotes show a discy structure in their Fourier components, and there are possible indications of spiral structure.  Thus we tentatively conclude that the host of SBS~$0846+513$ is a bulge + disc galaxy.  Either deeper ground-based observations with full adaptive optics or space-based observations with the {\it Hubble Space Telescope} or the {\it James Webb Space Telescope} would improve the resolution and the model of the galactic bulge, which is unresolved here.

\subsection{AGN Environment}\label{sec:environment}

We considered whether or not SBS~$0846+513$ has a companion, perhaps involved in a merger that could have triggered its active phase.  In Fig.~\ref{fig:field-annotated}, we see that Galaxy B is separated from SBS~$0846+513$ by a projected distance of 2.45 arcsec.  Their proximity has caused problems in the past.  In fact, SBS~$0846+513$ was once \citep{1989ApJS...69....1H} placed at a redshift of $z=1.86$, due to the contamination of its spectrum by Galaxy B, whose redshift was roughly modeled as 0.235 in prior papers.  At the resolution of the LBT, there is still a very slight overlap, and the possibility of a merger is interesting, given the role mergers sometimes play in triggering AGN.

From our spectroscopy (\S\ref{sec:spectroscopy}), we find the redshift of SBS~$0846+513$ is $z=0.585$, using H~$\beta$ and [OIII] lines, but the redshift of Galaxy B is 0.311, using H$\alpha$ and [NII] lines.  So the two galaxies are only visual companions, which are not interacting with each other.  

Some of the three features marked in Fig.~\ref{fig:host-residual} could be dwarf companions of the AGN.  If so, it would be consistent with the studies that show a majority of $\gamma$-NLS1 have signs of mergers or interactions \citep{2018rnls.confE..15K}.  The small clump to the east lies at a projected distance of 23 kpc (3.5 arcsec) from the AGN.  We have leaned towards interpreting the longer features to the northwest and southeast as arms of the AGN host, but it is possible they are also companions, if they are at the AGN redshift at all.  A 3-pixel Gaussian smoothing is applied to the image in the bottom row to show them better.

\subsection{Black Hole Mass}

Earlier mass estimates for SBS~$0846+513$ have come from a variety of methods.  \citet{2005ChJAA...5...41Z} puts its mass at $8.2 \times 10^6 \, \text{M}_\odot$ using the Broad-Line Region (BLR) H~$\beta$ linewidth, $5.2\times10^7 \, \text{M}_\odot$ using the $\text{M}_\mathrm{BH}$--$\sigma_*$ relation with the Narrow-Line Region [OIII] to get stellar motion in the bulge, or $4.3\times10^7 \, \text{M}_\odot$ using the $\text{M}_\mathrm{BH}$--$\text{M}_\mathrm{bulge}$ relation with the elliptical galaxy mass-to-light ratio and a host template spectrum.  \citet{2008ApJ...685..801Y} obtain $2.5\times10^7 \, \text{M}_\odot$ using the BLR H~$\beta$ linewidth, which is close to the $3.2\times10^7 \, \text{M}_\odot$ calculated by \citet{2015A&A...575A..13F}.  Using the estimates of black hole mass uncertainties by \citet{2004ASPC..311...69V}, both of these are consistent with the non-BLR masses of \citet{2005ChJAA...5...41Z}.  \citet{2011ApJS..194...45S} derive mass estimates for SBS~$0846+513$ based on {\it Sloan Digital Sky Survey} spectra, including H~$\beta$ and MGII-derived masses by a variety of calibrations.  These range from $5.6\times10^7$--$9.7\times10^7 \, \text{M}_\odot$.  These are slightly higher than the masses discussed above, but given their uncertainties, the low end of this range is consistent with the H~$\beta$ estimate of \citet{2005ChJAA...5...41Z}.

As mentioned in \S\ref{sec:intro}, it has been argued that the narrow widths of NLS1s' permitted lines might not be due to smaller black hole masses but to geometry--that these galaxies are viewed nearly pole-on to a disc-like BLR, minimizing the Doppler effects and leading to an underestimate of the black hole mass.  In gamma-ray emitting NLS1s, at least, we are more likely to have a nearly pole-on view of the black hole, since we must be looking along the radio jet.  But since we are able to resolve much of the host galaxy in the case of SBS~$0846+513$, we can derive measures of the black hole mass that are independent of the BLR flattening and orientation, using the correlations with bulge luminosity and radial profile.

Being derived from the host galaxy, the masses we calculate in Table~\ref{tab:bhmass} are independent of BLR geometry and viewing angle.  Using our preferred (bulge + disc) host morphology, the mass-luminosity relation of the bulge gives us the lowest of the four mass estimates.  At $5.0\times10^7 \, \text{M}_\odot$, it is comparable to the upper end of the range of masses from the prior work on this AGN.  The mass-luminosity relation assumes a classical bulge, and pseudobulges tend to have smaller black hole masses for their luminosity \citep{2012ApJ...754..146M}.  Since our bulge is effectively unresolved (the effective radius is narrower than the PSF), we don't know if it is classical or pseudo, and so the black hole mass could be lower.  On the other hand, if we accept the one-component model, then we count the luminosity of the entire host galaxy as the `bulge,' and we obtain our highest estimate, $1.6\times10^8 \, \text{M}_\odot$.  The S\'{e}rsic index methods applied to the one-component model give results that are close to this upper value, but because the bulge in the two-component model is so narrow, we don't try to derive a mass from its profile.  So the biggest variation in the mass estimates comes from adopting either the one or the two-component model.  In the end, the low end of our mass estimates (our preferred model) is consistent with previous results, while the high end is a factor of two or three times larger than these.  This indicates that the H~$\beta$-derived masses for this source are {\it not} substantially underestimated.

How does the mass of SBS~$0846+513$ compare with other NLS1 in general and gamma-ray emitters in particular?  Based on a sample selected by \citet{2017ApJS..229...39R}, \citet{2020CoSka..50..270B} report on their study of 3933 NLS1 whose H~$\beta$ lines fit a Gaussian or Lorentzian profile, using SDSS data extending out to a redshift of $z=0.8$.  Of these, none has a black hole mass greater than $10^8 \, \text{M}_\odot$.  Even our lowest mass estimate for SBS~$0846+513$ would make it more massive than all but 1.8 per cent (72 out of 3932) of the complete \citet{2020CoSka..50..270B} sample.  A better comparison would account for the variation in mass distribution with redshift.  Restricting ourselves to sources within a redshift range of $\pm0.1$ from SBS~$0846+513$ ($0.485 < z < 0.685$), we see that it is, at minimum, more massive than all but 4.5 per cent (35 out of 776) of NLS1 in that range.  Regarding gamma-ray emitting NLS1, the 10 other $\gamma$-NLS1 with measured masses \citep{2015A&A...575A..13F} range from $1.4\times10^7 \, \text{M}_\odot$ up to $1.15\times10^8 \, \text{M}_\odot$.  Our mass calculations for SBS~$0846+513$ puts it at least in the upper half of this group (more massive than 5 out of 10, at the low end), and at the high end, more massive than all of the others.

\citet{2018rnls.confE..15K} compiles mass estimates for two $\gamma$-NLS1 (1H~$0323+342$ and PKS~$1502+036$) whose black holes have been measured with several techniques, including host galaxy correlations (original host analyses by \citealt{2014ApJ...795...58L} and \citealt{2018MNRAS.478L..66D}, respectively), single-epoch H~$\beta$ spectroscopy, SED modeling, and others.  It is interesting to note that the host-galaxy-derived methods return much higher masses than the single-epoch H~$\beta$ spectra in those cases: by one order of magnitude for 1H~$0323+342$ and by two orders of magnitude for PKS~$1502+036$.  In fact, the host galaxy correlations lead to the highest mass estimates of {\it any} of the methods on the list.  Since single-epoch spectra are orientation-dependent while the host galaxy methods are not, one possibility is that the spectra are underestimating the black hole mass.  However, the H~$\beta$ results in these two cases are more consistent with mass results from several other methods, as we expect.  The consistency is good for 1H~$0323+342$; in the case of PKS~$1502+036$, the H~$\beta$ result returns the lowest mass of four methods across seven studies, but the greatest difference is between H~$\beta$ and the host galaxy result.  

On the other hand, the host galaxy methods could be overestimating the black hole mass.  \citet{2014ApJ...795...58L} find similar goodness-of-fit for the elliptical (bulge) and spiral (bulge + disc) models for 1H~$0323+342$'s host, and they choose the elliptical model for simplicity.  The bulge in the $J$-band spiral model is 1.5 mag fainter than in the elliptical model, and it has a S\'{e}rsic index of $n=0.88$, which is consistent with a pseudobulge.  For PKS~$1502+036$, \citet{2018MNRAS.478L..66D} find a lower $\chi^2/\nu$ for their spiral model than the elliptical model.  Choosing the spiral would make the bulge 0.4 mag fainter.  The S\'{e}rsic index is fixed at $n\equiv4$ in this case, so we cannot tell from their report if the bulge would be classical or a pseudobulge.  However, \citet{2020MNRAS.492.1450O} find a spiral host with a bulge whose S\'{e}rsic index is close to 1, indicating a pseudobulge and decreasing the black hole mass.  In both of these cases, then, using the spiral model to derive the black hole mass would be justifiable, and it would return a lower result.

It would be interesting to make more general comparisons between the masses of the known $\gamma$-NLS1 and the non-$\gamma$-emitters, but we must be cautious using even the H~$\beta$-derived masses for each group, because of different assumed values for the virial factor, $f$.  For example, the \citet{2020CoSka..50..270B} sample of NLS1 uses $f=3/4$, while the $\gamma$-NLS1 masses of \citet{2015A&A...575A..13F} are calculated using $f=3.85$.  Since $f$ is related, in part, to the inclination angle of the Broad-Line Region, and the $\gamma$-NLS1 must be viewed close to pole-on, then these different values may be appropriate.  But more work is needed, and it would be good to have a larger set of orientation-independent measures of black hole mass as a comparison.

It should be noted that there is a very strong statistical argument that the majority of Narrow-Line Seyfert 1s cannot have BH masses as high as Broad-Line Seyfert 1s (BLS1) and simply be underestimated due to their geometry.  As pointed out by \citet{2006AJ....132..531K}, if NLS1s were just the face-on analogues of BLS1s and narrower Balmer line widths because the Broad Line Region was flattened and viewed close to pole-on, then we should see a large fraction of beamed NLS1s.  In fact, more NLS1s would be seen as beamed than BLS1s, because they would be the population with more face-on sources.  Yet the opposite is the case.  Studies consistently find that the fraction of radio-loud sources is smaller in NLS1s.

So SBS~$0846+513$ has a massive black hole, larger than most NLS1, but consistent with estimates of the other gamma-ray emitting NLS1.  The H~$\beta$-derived masses for this source are not greatly underestimated.  

\subsection{Variability}

SBS~$0846+513$ shows strong variability at all wavelengths, from radio (e.g., \citealt{2015A&A...575A..55A}) to mid-IR (see \citealt{2015MNRAS.451.1795C}) to $\gamma$-rays (see \citealt{2015A&A...575A..13F}).  Comparing the photometry of SBS~$0846+513$ between our observations and the 2MASS Extended Source Catalog, we see the AGN nucleus has dimmed significantly since the 2MASS observation of 1998 December 23.  The 2MASS observations are not able to resolve the AGN and Galaxy B, so we subtract the light we measure for the host galaxies of these two and find the nucleus in 1998 had $J=15.9$.  So it has dimmed by 4.4 mag to the $J=20.3$ we observed on 2016 December 15.  SBS~$0846+513$ is known to have exceptional changes in the optical (e.g., the 4-magnitude variation in about a month recorded by \citealt{1979ApJ...230...68A}), so this level of variability is not unusual for this object and is well within the historical range.

\section{Conclusions}

We observed the Narrow-Line Seyfert 1 galaxy SBS~$0846+513$ in the $J$-band with the LUCI1 camera, aided by laser-guide ground-layer adaptive optics system, ARGOS, on the Large Binocular Telescope.  The galaxy has a significant gamma-ray emission, likely from a relativistic jet.  Very little is known about the host galaxies of the $\gamma$-NLS1 and how they fit into the broader population of normal NLS1.

The host galaxy is resolved.  There is a slight preference for the spiral model over the elliptical, and we see evidence for what may be spiral arms.  This and its discy isophotes argue for a bulge + disc host galaxy, which is consistent with most NLS1 at lower redshift, even though relativistic jets are more common in ellipticals.  Space-based observations from the {\it Hubble Space Telescope} or the {\it James Webb Space Telescope} would be useful to improve the resolution of the host, model the bulge, and look for signs of galaxy interactions.

Using correlations between black hole mass on the one hand and bulge luminosity and S\'{e}rsic index on the other, we find it harbors a central black hole with a mass likely around $5.0\times10^7 \, \text{M}_\odot$, but possibly as high as $1.6\times10^8 \, \text{M}_\odot$.  These measurements are independent of the viewing angle to the Broad Line Region and therefore avoid any biases due to BLR geometry, and they indicate that the H~$\beta$-derived masses for this object are not greatly underestimated.  The mass lies at the high end of the normal NLS1 range but is consistent with other $\gamma$-NLS1 masses.

\section*{Acknowledgements}

T. S. H. would like to thank Luigi Foschini for many useful discussions, especially on the interpretation of the mass.

S. A. acknowledges financial support from the Centre for Research \& Development in Mathematics and Applications (CIDMA) strategic project UID/MAT/04106/2019 and from Enabling Green E-science for the Square Kilometre Array Research Infrastructure (ENGAGESKA), POCI-01-0145-FEDER-022217, funded by the Programa Operacional Competitividade e Internacionaliza\c{c}\~{a}o (COMPETE 2020) and FCT, Portugal.

This paper is based on observations made with the Large Binocular Telscope and the Nordic Optical Telescope.  The Large Binocular Telescope is managed by the Large Binocular Telescope Observatory Collaboration, headquartered at the University of Arizona, in Tucson, Arizona, U.S.A.  The Nordic Optical Telescope is operated by the Nordic Optical Telescope Scientific Association at the Observatorio del Roque de los Muchachos, La Palma, Spain, of the Instituto de Astrofisica de Canarias.  Our research has made use of the NASA/IPAC Extragalactic Database (NED), which is funded by the National Aeronautics and Space Administration and operated by the California Institute of Technology.

\section*{Data Availability}

The data underlying this article will be shared on reasonable request to the corresponding author.




\bsp	
\label{lastpage}
\end{document}